\documentclass{Rinton-P9x6}
\usepackage{amsmath}
\usepackage{amssymb}
\usepackage{latexsym}
\usepackage{revsymb}

\newtheorem{defi}{Definition}
\newtheorem{lemma}[defi]{Lemma}
\newtheorem{thm}[defi]{Theorem}
\newtheorem{cor}[defi]{Corollary}
\newtheorem{rem}[defi]{Remark}
\newtheorem{prop}[defi]{Proposition}

\newcommand{\qed}{\hfill $\Box$}
\newcommand{\tr}{{\operatorname{Tr}}}
\newcommand{\id}{{\operatorname{id}}}
\newcommand{\supp}{{\operatorname{supp}\,}}

\newcommand{\bra}[1]{{\langle{#1}|}}
\newcommand{\ket}[1]{{|{#1}\rangle}}
\newcommand{\ketbra}[1]{{\ket{#1}\!\bra{#1}}}
\newcommand{\C}{{\mathbb{C}}}

\newcommand{\N}{{\mathbb{N}}}

\newcommand{\fset}[1]{{\mathcal{#1}}}

\newcommand{\1}{{\openone}}

\newlength{\blank}
\newlength{\equalsign}
\newenvironment{beweis}[1][{\hspace{-\blank}}]{{\noindent\emph{Proof~{#1}.\ }}}{\hfill $\Box$\vskip 0.5\baselineskip}

\begin{document}

\title{Quantum and Classical Message\protect\\ 
        Identification via Quantum Channels}

\author{Andreas Winter}
\address{School of Mathematics, University of Bristol,\\
University Walk, Bristol BS8 1TW, U.K.\\
Email: {\tt a.j.winter@bris.ac.uk}}

\maketitle

\begin{center}
  {\bf (Dedicated to Alexander~S.~Holevo on his 60th birthday)}
\end{center}
\bigskip

\abstracts{We discuss
  concepts of message identification in the sense of
  Ahlswede and Dueck via general quantum channels, extending
  investigations for classical channels, initial work for
  classical--quantum (cq) channels and ``quantum fingerprinting''.
  We show that the identification capacity of a discrete memoryless
  quantum channel for classical information can be larger than that
  for transmission; this is in contrast to all previously considered
  models,
% (including so--called ``simultaneous'' identification)
  where it turns out to equal the common randomness capacity
  (equals transmission capacity in our case): in particular, for a
  noiseless qubit, we show the identification capacity to be $2$, while
  transmission and common randomness capacity are $1$.
%  This is an effect connected to state
%  space geometry rather than dense coding as demonstrated
%  by quantum--classical (qc) channels whose identification
%  capacity equals their transmission capacity but which
%  typically have larger entanglement--assisted capacities.
  Then we turn to a natural concept of identification of \emph{quantum}
  messages (i.e.~a notion of ``fingerprint'' for quantum states).
  This is much closer to quantum information transmission than its classical
  counterpart (for one thing, the code length grows only exponentially,
  compared to double exponentially for classical identification).
  Indeed, we show how the problem exhibits a nice connection
  to \emph{visible} quantum coding.
  Astonishingly, for the noiseless qubit channel this capacity
  turns out to be $2$: in other words, one can compress two qubits into one
  and this is optimal. In general however, we conjecture quantum identification
  capacity to be different from classical identification capacity.}

\section{Introduction}
\label{sec:intro}
The theory of identification of messages via (noisy) channels was initiated
by Ahlswede and Dueck,\cite{AD1,AD2} and has resulted in interesting
developments: it has a strong connection to the
theory of common randomness,\cite{AC} and it has triggered
the theory of ``approximation of output statistics''.\cite{Han:Verdu,steinberg}
Put briefly, whereas in Shannon's famed theory of communication\cite{Shannon}
the object is to \emph{transmit} a message over a channel, described by a
stochastic map $W:\fset{X}\longrightarrow\fset{Y}$, by block coding of
length $n$, for identification the receiver should only be able to
answer questions ``Is the message $i$ sent equal to one $j$ that I have in
mind?''. It turns out that this achieves much larger codes than transmission:
while the latter has an exponentially growing (in $n$)
optimal message set, the former allows for \emph{double exponential growth}.
Rabin and Yao\cite{Rabin:Yao} --- see also Kushilevitz and Nisan,\cite{Kushilevitz:Nisan}
Example 3.6 and the Bibliographic Notes --- have shown
that the randomised communication complexity of the equality
function of $n$--bit strings is $O(\log n)$ bits.
The achievement of Ahlswede and Dueck\cite{AD1} was to
determine the constant in the exponent: it is the Shannon
capacity of the channel --- but can be larger in models with
feedback.\cite{AD2}
However, in this paper we will only consider discrete memoryless
(classical and quantum) channels
with no feedback or other helpers, and with an i.i.d.~time structure.
\par
Formally, consider a discrete memoryless quantum channel\cite{Holevo77}
$$T:{\cal A}_1\longrightarrow{\cal A}_2,$$
modelled as a completely positive, trace preserving
map between C${}^*$--algebras ${\cal A}_1$, ${\cal A}_2$
which in this paper are always finite dimensional.
It is well--known that a finite C${}^*$--algebra
${\cal A}$ is isomorphic to a direct sum of full matrix
algebras, and we can think of it as the subalgebra
of the operator algebra ${\cal B}({\cal H})$ on a finite
dimensional Hilbert space ${\cal H}$, of operators commuting
with some selfadjoint operator $A$:
$${\cal A} = \{X: XA=AX\} = \bigoplus_{i=1}^r {\cal B}({\cal H}_i),$$
with the eigenspaces ${\cal H}_i$ of $A$.
The states on ${\cal A}$ we identify with the semidefinite
operators in ${\cal A}$ of trace $1$, the set of which we denote ${\cal S}({\cal A})$.
Following Holevo\cite{Holevo77} we call $T$ a \emph{cq--channel}
if ${\cal A}_1$ is commutative ($=$classical), and a
\emph{qc--channel} if ${\cal A}_2$ is commutative: in the first
case the action of $T$ is determined by its images $W_x=T(x)$ on
the (finitely many) minimal idempotents $x$, in the second case
this applies to the adjoint map $T^*:{\cal A}_2\longrightarrow{\cal A}_1$,
and the images $M_y=T^*(y)$ of the minimal idempotents $y$
form a positive operator valued measure (POVM).
The channel is classical if both ${\cal A}_1$ and ${\cal A}_2$
are commutative.
\par
We shall consider block coding for $n$ copies of $T$:
an \emph{$(n,\lambda)$--code} for this channel
is a collection $\{(\pi_i,D_i):i=1,\ldots,M\}$
of (without loss of generality: pure) states $\pi_i$ on ${\cal A}_1^{\otimes n}$,
and positive operators $D_i\in{\cal A}_2^{\otimes n}$ which sum to $\1$
(i.e., a POVM), such that
\begin{equation}
  \label{eq:c:general:quantum}
  \forall i\quad \tr\bigl(T^{\otimes n}(\pi_i) D_i\bigr) \geq 1-\lambda.
\end{equation}
The maximum such $M$ will be denoted $M(n,\lambda)$. If all the $\pi_i$
are product states with respect to ${\cal A}_1^{\otimes n}$, the code is
called \emph{separable}, and the corresponding maximal $M$ is denoted
$M_1(n,\lambda)$.
The knowledge about these quantities is summarised in the
HSW--theorem ($H(\rho)=-\tr\rho\log\rho$ is the von Neumann
entropy):
\begin{thm}{\bf (Holevo,\cite{Holevo73,Holevo79,Holevo98}
            Schumacher, Westmoreland,\cite{SW97}
            Ogawa, Nagaoka,\cite{Ogawa:Nagaoka}
            Winter\cite{Winter99,winter:diss})}
  \label{thm:holevo:strong}
  For all $0<\lambda<1$, one has the coding theorem and \emph{strong converse},
  $$\lim_{n\rightarrow\infty} \frac{1}{n}\log M_1(n,\lambda) = \chi(T),$$
  with the \emph{Holevo capacity} of the channel
  $$\chi(T)=\max_{\{p_i,\pi_i\}}
             \left\{ H\left(\sum_i p_i T(\pi_i)\right)
                      - \sum_i p_i H\bigl(T(\pi_i)\bigr) \right\}.$$
  For general codings, the capacity and the \emph{weak converse} are given by
  \begin{equation*}\begin{split}
    C(T) &= \inf_{\lambda>0}\liminf_{n\rightarrow\infty} \frac{1}{n}\log M(n,\lambda) \\
         &= \inf_{\lambda>0}\limsup_{n\rightarrow\infty} \frac{1}{n}\log M(n,\lambda)
          =\lim_{n\rightarrow\infty} \frac{1}{n}\chi(T^{\otimes n}).
  \end{split}\end{equation*}
  (Here and elsewhere in the paper $\log$ and $\exp$ are to basis $2$.)
  \qed
\end{thm}
It is widely conjectured that $\chi$ is additive with respect to
tensor products, which would imply $C(T)=\chi(T)$. Furthermore, it is conjectured
that the strong converse also holds for $M(n,\lambda)$ (which is known
for cq--channels since there $M(n,\lambda)=M_1(n,\lambda)$).
\par
Following L\"ober\cite{Loeber} and generalising the
classical definition\cite{AD1} an
\emph{$(n,\lambda_1,\lambda_2)$--ID code} for $T$ is a
collection of pairs $\{(\rho_i,D_i):i=1,\ldots,N\}$ of
states $\rho_i$ on ${\cal A}_1^{\otimes n}$
and operators $0\leq D_i\leq \1$ in ${\cal A}_2^{\otimes n}$, such that
\begin{equation}
  \label{eq:c-ID:general:quantum}
  \begin{aligned}
    \forall i       &\quad \tr\bigl(T^{\otimes n}(\rho_i) D_i\bigr)\geq 1-\lambda_1, \\
    \forall i\neq j &\quad \tr\bigl(T^{\otimes n}(\rho_i) D_j\bigr)\leq \lambda_2.
  \end{aligned}
\end{equation}
The bounds $\lambda_1$ and $\lambda_2$ are called \emph{error probabilities
of first} and \emph{second kind}, respectively, since the problem is
a coding variant of hypothesis testing.
For the identity channel $\id_{\cal A}$ of a system ${\cal A}$
(Kuperberg\cite{kuperberg} calls this a \emph{hybrid quantum memory}),
we call an ID code also a code \emph{on the algebra ${\cal A}^{\otimes n}$},
or more generally on an algebra $\widetilde{{\cal A}}$ if the
$\rho_i$ and $D_i$ are elements of $\widetilde{{\cal A}}$.
\par
The code is called \emph{simultaneous} if the binary observables
(POVMs) $(D_i,\1-D_i)$ are all co--existent, in the sense of
Ludwig,\cite{ludwig:axiomatic} i.e. if there exists a POVM
$(E_k)_{k=1,\ldots,K}$ such that for all $i$ a set
$\fset{D}_i\subset\{1,\ldots,K\}$ can be found with
\begin{equation}
  \label{eq:sim}
  D_i=\sum_{k\in\fset{D}_i} E_k.
\end{equation}
Denote the maximum $N$ such that an $(n,\lambda_1,\lambda_2)$--ID code
(a simultaneous $(n,\lambda_1,\lambda_2)$--ID code)
of length $N$ exists by $N(n,\lambda_1,\lambda_2)$
($N_{\rm sim}(n,\lambda_1,\lambda_2)$). Clearly,
$N_{\rm sim}(n,\lambda_1,\lambda_2) \leq N(n,\lambda_1,\lambda_2)$.
\begin{thm}[L\"ober\cite{Loeber}]
  \label{thm:simID}
  For all $\lambda_1,\lambda_2>0$,
  $$ \liminf_{n\rightarrow\infty} \frac{1}{n}\log\log N_{\rm sim}(n,\lambda_1,\lambda_2)
                                                                                \geq C(T),$$
  with the Holevo (transmission) capacity $C(T)$ of the channel.
  If the channel $T$ satisfies the strong converse and in addition
  the technical condition that on block length $n$ its
  input states may be restricted to an alphabet of size $2^{2^{o(n)}}$,
  then for $0<\lambda_1,\lambda_2$, $\lambda_1+\lambda_2<1$,
  \begin{equation*}
    \lim_{n\rightarrow\infty} \frac{1}{n}\log\log N_{\rm sim}(n,\lambda_1,\lambda_2)=C(T).
  \end{equation*}
  \qed
\end{thm}
The first part is proved by concatenating a transmission code
for $T$ with the following construction:
\begin{prop}[Ahlswede, Dueck\cite{AD1}]
  \label{prop:AD-construction}
  Let ${\cal M}$ be a set of cardinality $M$, $\lambda>0$ and $\epsilon$ such that
  $\lambda\log\left(\frac{1}{\epsilon}-1\right)>2$. Then there exist
  $N\geq 2^{\lfloor\epsilon M\rfloor}/M$ subsets ${\cal M}_i\subset{\cal M}$
  of cardinality $\lfloor\epsilon M\rfloor$, such that
  $$\forall i\neq j\quad |{\cal M}_i\cap{\cal M}_j|\leq \lambda\lfloor\epsilon M\rfloor.$$
  In other words, the pairs $\{(P_i,{\cal M}_i):i=1,\ldots,N\}$, with
  the uniform distribution $P_i$ on ${\cal M}_i$, form an ID code
  with error probability of first kind $0$, and of second kind $\lambda$.
  \qed
\end{prop}
The technical condition in theorem~\ref{thm:simID} is true for example
for cq--channels ($2^{O(n)}$ input strings) but not for the ideal
qubit channel $\id_{\C^2}$: to approximate every input state on $n$
qubits requires $2^{2^{const.n}}$ pure states. It is not known
if the condition is necessary for the conclusion of the theorem
but our construction following below in section~\ref{sec:c:ID}
certainly violates it.
\par
Regarding the necessity of the simultaneity condition, we have
\begin{thm}[Ahlswede, Winter\cite{AW}]
  \label{thm:cq-ID}
  For a cq--channel $T$ and $0<\lambda_1,\lambda_2$, $\lambda_1+\lambda_2<1$,
  \begin{equation*}
    \lim_{n\rightarrow\infty} \frac{1}{n}\log\log N(n,\lambda_1,\lambda_2)
                                                                 =C(T)=\chi(T).
  \end{equation*}
  The same holds for general channels and codes of separable
  states, as this restricts us effectively to a cq--channel.
  \qed
\end{thm}
\par\medskip
This left open, however, the understanding of the precise role
of the simultaneity condition in general.
The present work will push this question further a bit:
%
%The present work will illuminate this question
%by showing in section~\ref{sec:c:ID} first
%that also for qc--channels one can drop simultaneity
%without changing the capacity, and that in general the simultaneous
%identification capacity equals the transmission capacity.
%Secondly, however, it can make a tremendous difference for general quantum channels:
we determine the capacity to be $2$ for the noiseless qubit channel
(theorem~\ref{thm:ID:quantum:identity}),
and give a formula for general hybrid quantum memories
(corollary~\ref{cor:hybrid}), based on a new construction
to extend an identification code.
We furthermore conjecture
the capacity to be different for \emph{simultaneous}
identification: specifically, that it is $1$ for the
noiseless qubit.
\par
While all this is concerned with identification of classical messages,
the second part of the paper (section~\ref{sec:q:ID}) will study
concepts of \emph{quantum message} identification.
These can be related to \emph{visible} codings and decodings
for quantum channels. Roughly, quantum message identification aims at
encoding the pure states $\ket{\phi}$ of a Hilbert space, such that 
a quantum mechanical test of ``Is it $\ket{\theta}$?'' may performed
on the output: a binary measurement with distribution close to
$\bigl(|\langle\theta\ket{\phi}|^2,1-|\langle\theta\ket{\phi}|^2\bigr)$.
I.e., the fidelity test can be simulated, which has been argued to be
the quantum analogue of the identity predicate.\cite{Winter01}
Our findings are in sharp contrast to the case of classical message identification:
we do not find the doubly exponential growth of code length (here: dimension)
with the block length, but only exponential growth, like in transmission.
Still, we find a capacity different from the
quantum transmission capacity: it is $2$ for the noiseless qubit
channel (theorem~\ref{thm:2for1}).
We conclude with a discussion of our results which puts them in
their right context, and highlight open questions and conjectures.

\section{Classical message identification}
\label{sec:c:ID}
The results of L\"ober\cite{Loeber} and of Ahlswede and the
author\cite{AW} seem to indicate that the (simultaneous
and non--simultaneous) identification capacity
of a memoryless quantum channel is equal to its
transmission capacity.
\par
We now show that for a general channel --- in fact 
we may take a noiseless qubit ---
the (non--simultaneous) identification
capacity can exceed the transmission capacity.
We begin with a result about \emph{pure state} ID codes:
\begin{prop}[``Quantum fingerprinting''\cite{BCWdW}]
  \label{prop:many:pure:states}
  On ${\cal B}(\C^d)$ there exists an
  ID code $\{(\psi_i,\psi_i):i=1,\ldots,N\}$ of
  $N\geq 2^{\lfloor\epsilon d\rfloor}/d$ pure states
  $\psi_i=\ketbra{\psi_i}$,
  with error of first kind $0$ and of second kind $\lambda$,
  with $\lambda\log\left(\frac{1}{\epsilon}-1\right)>4$.
\end{prop}
\begin{beweis}
  We could simply take the construction of Buhrman \emph{et al.},\cite{BCWdW}
  but we prefer to reduce it to proposition~\ref{prop:AD-construction}:
  \par
  Take the probability
  distributions $P_i$ on $\{1,\ldots,d\}$ of an ID code
  $\{(P_i,{\cal M_i})\}$ such as proposition~\ref{prop:AD-construction}
  (with error probability of second kind $\lambda/2$) and define the states
  $\ket{\psi_i}=\sum_{k=1}^d \sqrt{P(k)}\ket{k}$. Note that
  defining $D_i=\ketbra{\psi_i}$ will make the error probability
  of first kind equal to $0$. For the error of second kind, note
  that it is (for $i\neq j$)
  $$|\bra{\psi_i}\psi_j\rangle|^2 = \left(\sum_k \sqrt{P_i(k)P_j(k)}\right)^2.$$
  Using well--known relations between distinguishability measures
  of distributions (see e.g.~Fuchs and van de Graaf\cite{fuchs:vandegraaf})
  this is upper bounded by $\lambda$.
\end{beweis}
Using mixed states, one can increase $N$ dramatically:
\begin{prop}
  \label{prop:many:mixed:states}
  For every $0<\lambda<1$ and $\epsilon>0$ such that
  $\lambda\log\left(\frac{1}{\epsilon}-1\right)>8$, there exists on the quantum system
  ${\cal B}(\C^d)$ an ID code with
  $$N \geq \frac{1}{K(\lambda)d^2}
              \exp\left(\bigl\lfloor\epsilon K(\lambda)d^2\bigr\rfloor\right)$$
  messages, with error probability of first kind equal to $0$ and error
  probability of second kind bounded by $\lambda$.
  The constant $K(\lambda)$ may be chosen
  $$K(\lambda)=\frac{(\lambda/100)^4}{4\log(100/\lambda)}.$$
\end{prop}
\begin{beweis}
  We concatenate quantum fingerprinting, proposition~\ref{prop:many:pure:states},
  with the main result of the next section, proposition~\ref{prop:2for1}:
  according to it one can encode the pure states on $\C^S$,
  $S=\lfloor K(\lambda)d^2\rfloor$, into
  ${\cal B}(\C^d)$ such that the measurements $(\pi,\1-\pi)$
  can be implemented with accuracy $\lambda/2$ in the individual
  output probabilities.
  We simply encode the states from a pure state ID code with error of
  second kind $\lambda/2$. Looking at the proof of proposition~\ref{prop:2for1}
  we see that we can assume that the fingerprinting states are part
  of the net of states (see the following lemma) on which the error of first
  kind does not increase --- it stays $0$.
\end{beweis}
Using the following lemma we can show that the cardinalities
of the codes in propositions~\ref{prop:many:pure:states}
and~\ref{prop:many:mixed:states} are asymptotically optimal.
\begin{lemma}[Bennett \emph{et al.},\cite{rand} lemma 4]
  \label{lemma:net}
  For $\epsilon>0$, there is a set $\fset{M}$ of pure states
  of pure states in $d$--dimensional Hilbert space with
  $|\fset{M}|\leq (5/\epsilon)^{2d}$, such that
  for all pure states $\ket{\varphi}$ there is 
  $\ket{\tilde{\varphi}}\in\fset{M}$ with
  $\bigl\| \ketbra{\varphi}-\ketbra{\tilde{\varphi}} \bigr\|_1\leq\epsilon$.
  (We call such a set an \emph{$\epsilon$--net}.)
  \qed
\end{lemma}
\begin{prop}
  \label{prop:optimality}
  Let $\{(\rho_i,D_i):i=1,\ldots,N\}$ be an ID code on ${\cal B}(\C^d)$
  with error probabilities $\lambda_1,\lambda_2$ of first and second kind, respectively,
  with $\lambda_1+\lambda_2<1$. Then,
  \begin{align*}
    \text{if all }\rho_i\text{ are pure},\quad &
                  N\leq \left(\frac{5}{1-\lambda_1-\lambda_2}\right)^{2d}, \\
    \text{for general }\rho_i,\quad            &
                  N\leq \left(\frac{5}{1-\lambda_1-\lambda_2}\right)^{2d^2}.
  \end{align*}
\end{prop}
\begin{beweis}
  The key insight is that for $i\neq j$,
  $\frac{1}{2}\|\rho_i-\rho_j\|_1 \geq 1-\lambda_1-\lambda_2$,
  because we have an ID code.
  \par
  In the first case of pure $\rho_i$, fix an $\epsilon$--net ${\cal M}$
  in the pure states, with $\epsilon<1-\lambda_1-\lambda_2$,
  according to lemma~\ref{lemma:net}. This net decomposes
  the set of pure states into (Voronoi) cells of radius $\leq\epsilon$,
  so no two $\rho_i$ can be in the same cell. Hence
  $N\leq \left(\frac{5}{\epsilon}\right)^{2d}$, and as $\epsilon$
  was arbitrary we obtain the first upper bound.
  \par
  For the general case pick an $\epsilon$--net in $\C^d\otimes\C^d$,
  according to lemma~\ref{lemma:net}. Since every state on $\C^d$
  has a purification on $\C^d\otimes\C^d$ and the trace distance
  is monotonic under partial traces, we obtain an $\epsilon$--net
  in ${\cal S}(\C^d)$ of cardinality
  $\left(\frac{5}{\epsilon}\right)^{2d^2}$. From
  here we argue as in the pure state case.
\end{beweis}
We can now determine the identification capacity of the noiseless qubit:
\begin{thm}
  \label{thm:ID:quantum:identity}
  For the ideal qubit channel $\id_{\C^2}$ and $0<\lambda_1,\lambda_2$,
  $\lambda_1+\lambda_2<1$,
  $$\lim_{n\rightarrow\infty} \frac{1}{n}\log\log N(n,\lambda_1,\lambda_2)
                                                 =C_{\rm ID}(\id_{\C^2})=2.$$
\end{thm}
\begin{beweis}
  Immediate by putting together
  propositions~\ref{prop:many:mixed:states}
  and~\ref{prop:optimality}, for $d=2^n$.
\end{beweis}
\par
This shows that for a general channel $T$, the (classical message)
identification capacity is at least the double of its quantum transmission
capacity $Q(T)$ (i.e., the capacity of the channel to transmit
\emph{quantum} information\cite{bennett:shor,temaconvariazioni}).
Of course, by the general construction from proposition~\ref{prop:AD-construction}
it is also at least as large as its classical capacity $C(T)$.
In fact, we can show a bit more, with the help of the
following generalisation of the constructions of
Ahlswede and Dueck:\cite{AD1,AD2}
\begin{prop}
  \label{prop:blowup-code}
  Let $\{(\rho_i,D_i):i=1,\ldots,N\}$
  be an ID code on ${\cal A}$ with error probabilities
  $\lambda_1,\lambda_2$ of first and second kind, respectively,
  and let ${\cal C}$ be a classical system (commutative algebra)
  of dimension $M$.
  Then, for $\epsilon>0$, there exists an ID code
  $\{(\sigma_f,\widetilde{D}_f):f=1,\ldots,N'\}$
  on ${\cal A}\otimes{\cal C}$ with error probabilities
  $\lambda_1,\lambda_2+\epsilon$ of first and second kind, respectively,
  and $N'\geq\left(\frac{1}{2}N^\epsilon\right)^M$.
\end{prop}
\begin{beweis}
  Denote the minimal idempotents of ${\cal C}$ as $[k]$,
  $k=1,\ldots,M$.
  The new code elements and decoding operators
  will be constructed iteratively via
  (random) functions $f:\{1,\ldots,M\}\rightarrow\{1,\ldots,N\}$:
  \begin{equation*}
    \sigma_f        = \frac{1}{M}\sum_k \rho_{f(k)}\otimes[k], \quad
    \widetilde{D}_f =            \sum_k D_{f(k)}   \otimes[k].
  \end{equation*}
  Clearly, the error probability of first type is $\lambda_1$.
  Let a maximal code of this sort, with error probability
  of second kind $\lambda_2+\epsilon$, be already constructed,
  with functions $f_1,\ldots,f_{N'}$ as above.
  Pick the new function $f$ randomly, i.e.~all values $f(k)$ are
  uniform on $\{1,\ldots,N\}$ and independent, and construct
  $\sigma_f$ and $\widetilde{D}_f$ according to our prescription.
  \par
  It is easy to check that the error probability of second kind
  is bounded as
  \begin{equation}
    \label{eq:secondtype}
    \tr\bigl( \sigma_f\widetilde{D}_{f_j} \bigr),
       \tr\bigl( \sigma_{f_j}\widetilde{D}_f \bigr)
          \leq \lambda_2+\frac{1}{M}\bigl|\{k:f(k)=f_j(k)\}\bigr|.
  \end{equation}
  Fixing $j$ for the moment, we introduce the Bernoulli variables
  $X_k$ with $X_k=1$ if $f(k)=f_j(k)$ and $0$ otherwise. These
  are evidently independent and all have expectation $1/N$.
  By Sanov's theorem\cite{Dembo:Zeitouni} (with the binary relative
  entropy $D(p\|q)=p\log\frac{p}{q}+(1-p)\log\frac{1-p}{1-q}$),
  \begin{equation*}
    \Pr\left\{ \frac{1}{M}\sum_k X_k > \epsilon \right\}
                  \leq \exp\bigl( -M\,D(\epsilon\|1/N) \bigr)
                  \leq \exp\bigl( -M(\epsilon\log N - 1) \bigr).
  \end{equation*}
  With the union bound and using eq.~(\ref{eq:secondtype}), we can bound
  the probability that one of the error probabilities of second
  kind exceeds $\lambda_2+\epsilon$, by
  $$N'\exp\bigl( -M(\epsilon\log N - 1) \bigr);$$ 
  if this is less than $1$ we can hence enlarge our code, contradicting
  the maximality assumption. This implies the lower bound on $N'$.
\end{beweis}
\begin{rem}
  \label{rem:CR-assistance}
  We can make a connection to the second construction of Ahlswede and
  Dueck\cite{AD2} based on common randomness, observing that the distribution
  on $\{1,\ldots,M\}$ is always uniform; in other words, the code can be
  used with equal effect if the uniformly distributed $k\in\{1,\ldots,M\}$
  is known both to the sender and the receiver. Ahlswede and Dueck
  showed how to build an ID code of rate $R$, using negligible communication, 
  from rate $R$ of common randomness;
  proposition~\ref{prop:blowup-code} shows more generally
  how to \emph{increase} the rate of any given ID code by $R$.
  \par
  This construction (see also how we use it in the proof of
  theorem~\ref{thm:2Q+C} below) shows in generality that any
  additional classical capacity besides an ID code increases
  the identification capacity; hence, an optimal ID code for a channel
  does not allow for any remaining transmission rate.
  \par
  In this connection we can ask the interesting question if the analogue
  to theorem~\ref{thm:ID:quantum:identity}
  of the above--mentioned result by Ahlswede and Dueck\cite{AD2}
  holds with prior shared entanglement: do $nE$ ebits and negligible communication
  yield an identification rate of $2E$?
\end{rem}
\begin{thm}
  \label{thm:2Q+C}
  If the channel $T$ simultaneously transmits quantum
  information at rate $Q$ and classical information at rate $R$,
  then $C_{\rm ID}(T) \geq 2Q+R$.
  \par
  The joint quantum--classical capacity region was recently determined
  by Devetak and Shor,\cite{DS:qc} giving a lower bound of
  $$C_{\rm ID}(T) \geq \lim_{n\rightarrow\infty}
                        \frac{1}{n}\max_{\sigma^{XAB}} \bigl\{2 I_c(A\rangle BX)+I(X;B)\bigr\},$$
  where the maximisation is over all states
  $\sigma^{XAB}=\sum_x p_x\ketbra{x}^X\otimes\sigma_x^{AB}$,
  with $\sigma_x^{AB}=(\id\otimes T^{\otimes n})\phi_x$ for some
  bipartite pure states $\phi_x$.
  In the formula, $I(X;B)=H(\sigma^B)-\sum_x p_x H(\sigma_x^B)$
  is the familiar Holevo quantity,\cite{Holevo73}
  and $I_c(A\rangle BX)=H(\sigma^{XB})-H(\sigma^{XAB})$
  is the coherent information.\cite{schumacher96}
\end{thm}
\begin{beweis}
  For sufficient block length $n$, the channel can transmit a Hilbert
  space of dimension $d=2^{n(Q-\epsilon)}$ and simultaneously
  $K=2^{n(R-\epsilon)}$ messages (both with fidelity loss/error bounded
  by $\epsilon$).
  \par
  Using an ID code for $\C^d$ according to
  proposition~\ref{prop:many:mixed:states} and concatenating it with
  the construction of proposition~\ref{prop:blowup-code}
  we obtain an $(n,\lambda,\lambda)$--ID code of $N'\geq 2^{2^{n(2Q+R-\lambda)}}$
  messages, with $\lambda\rightarrow 0$ as
  $n\rightarrow\infty$ and $\epsilon\rightarrow 0$.
\end{beweis}
\begin{cor}
  \label{cor:hybrid}
  For a hybrid quantum memory ${\cal A}=\bigoplus_{i=1}^r {{\cal B}}(\C^{d_i})$,
  $$C_{\rm ID}(\id_{{\cal A}})
      =\max_{(p_1,\ldots,p_r)} \left\{ 2\sum_i p_i\log d_i + H(p) \right\}
      =\log \left( \sum_i d_i^2 \right).$$
\end{cor}
\begin{beweis}
  Denoting the maximum in the theorem by $\Gamma$,
  even the strong converse holds: for $\lambda_1+\lambda_2<1$,
  $$\lim_{n\rightarrow\infty} \frac{1}{n}\log\log N(n,\lambda_1,\lambda_2)
                                                       = \Gamma.$$
  First, the lower bound follows directly from theorem~\ref{thm:2Q+C}.
  \par
  The strong converse is proved by building an $\epsilon$--net for
  ${\cal S}({\cal A}^{\otimes n})$, and then arguing as in
  proposition~\ref{prop:optimality}; in detail, we first observe
  $${\cal A}^{\otimes n} \simeq \bigoplus_{P\ n\text{--type}}
                                     {\cal B}\left(\C^{2^{n\sum_i P_i\log d_i}}\right)
                                                            \otimes{\cal C}_{T(P)},$$
  where the \emph{$n$--types} are distributions on $\{1,\ldots,r\}$ with
  probability values $P_i\in\N/n$ and ${\cal C}_{T(P)}$ is a commutative algebra
  of dimension $T(P)\leq 2^{nH(P)}$, the number of sequences of type
  (i.e.~empirical distribution) $P$.
  The states in this algebra are obtained from states on
  $${\cal B}\left(\bigoplus_P \C^{2^{n\sum_i P_i\log d_i}} \otimes \C^{2^{n\sum_i P_i\log d_i}}
                                                              \otimes \C^{T(P)} \right)$$
  by dephasing different $P$ (this is a conditional expectation), tracing
  out the first tensor factor in each product, and subjecting the third factor
  to a complete von Neumann measurement. All this can be described by a
  quantum channel $R:{\cal B}(\C^N)\longrightarrow{\cal A}^{\otimes n}$,
  with $N\leq (n+1)^r 2^{n\Gamma}$, since there are $\leq (n+1)^r$ many
  $n$--types. Now we can invoke lemma~\ref{lemma:net} to construct an
  $\epsilon$--net for ${\cal S}(\C^N)$ which by virtue of $R$ gives an
  $\epsilon$--net for ${\cal S}({\cal A}^{\otimes n})$.
\end{beweis}
\begin{rem}
  \label{rem:simID}
  Now that we see that the identification capacity can be larger than
  the capacity for transmission, we may want to re--examine the significance
  of the condition of simultaneity: it effectively replaces the quantum
  channel by a classical one, with, on block length $n$, a doubly
  exponentially large net of input states, and classical output
  which is postprocessed for identification.
  \par
  Nevertheless, the single measurement may have the effect of likening
  the simultaneous model to transmission, and in particular
  we conjecture that for the ideal qubit channel the capacity is
  only $1$: with $\lambda_1+\lambda_2<1$,
  $$C_{\rm sim-ID}(\id_{\C^2})
       = \lim_{n\rightarrow\infty}\frac{1}{n}\log\log N_{\rm sim}(n,\lambda_1,\lambda_2)
       = 1.$$
  A justification for this may come from work on channel
  simulations:\cite{BDHSW} it can be shown that any measurement
  on $n$ qubits can, for all purposes of the output, be well
  approximated by a convex combination of measurements
  each of which has only $N=2^{n+o(n)}$ many outcomes. The convex
  combination represents randomisation at the output and while
  in general an ID code cannot be derandomised at the output,
  the quantum situation may be sufficiently special to allow
  such a conclusion. If that works, we would be done, since
  the output probabilities of a measurement of $N$ outcomes
  permits a net of $2^O(N)$ points.
\end{rem}

\section{Quantum message identification}
\label{sec:q:ID}
All the previous theory concerned identification of \emph{classical} messages,
though via quantum channels, and is thus analogous to classical information
transmission via quantum channels.\cite{Holevo98,SW97} But just as there is
also a theory of coding \emph{quantum information} for quantum
channels (see the review by Bennett and Shor\cite{bennett:shor}),
we can discuss the possibilities
of \emph{quantum message identification}.
\par
The following concepts appear rather natural: for the quantum
channel $T:{\cal A}_1\longrightarrow{\cal A}_2$
we define an \emph{$(n,\lambda)$--quantum--ID code} to be a pair
$(\varepsilon,D)$ of maps, such that
\begin{equation}
  \label{eq:quantum:encoder}
  \varepsilon:{\cal B}({\cal M})\longrightarrow{\cal A}_1^{\otimes n}
\end{equation}
is a quantum channel and
\begin{equation}
  \label{eq:quantum:decoder}
  D:{\cal P}({\cal M})\longrightarrow{\cal A}_2^{\otimes n}
\end{equation}
maps pure states $\tau=\ketbra{\theta}$ to operators $0\leq D_\tau\leq\1$,
with the properties that for all pure states
$\pi=\ketbra{\phi}$, $\tau=\ketbra{\theta}$ on ${\cal B}({\cal M})$:
\begin{equation}
  \label{eq:quantum:ID}
  \left|
     \tr(\pi\tau)
     -\tr\big(\left((T^{\otimes n}\varepsilon)\pi\right)D_\tau\big)
  \right|                                                      \leq\lambda/2.
\end{equation}
This means: the encoding is such that for the pure states
$\pi$ and $\tau$ a test (the binary POVM $(D_\tau,\1-D_\tau)$)
may be performed on the output signal $\bigl(T^{\otimes n}\varepsilon\bigr)\pi$
which is stochastically almost equivalent to the 
test $(\tau,\1-\tau)$ on the original state $\pi$.
Why should we regard this as ``quantum message identification''? The
conception that we ought to make a (highly error--free) decision if the
state $\pi$ equals $\tau$ we dismiss as unphysical: at the output we cannot
expect higher distinguishability than at the input. For the input, however, the
quantum mechanical version of the test ``Is $\pi$ equal to $\tau$ or not?''
has been argued\cite{Winter01} to be just $(\tau,\1-\tau)$. Also, restricted
to a subset of mutually orthogonal (or almost orthogonal) states on ${\cal M}$,
i.e.~for classical messages, the concept reduces to the classical
message identification codes discussed in section~\ref{sec:c:ID}.
\par
Now define $L(n,\lambda)$ to be the largest $\dim{\cal M}$ such that an
$(n,\lambda)$--quantum--ID code exists. For the identity channel on
some system $\widetilde{\cal A}$ we adopt the convention to speak of 
\emph{codes on $\widetilde{\cal A}$}, similar to the classical case;
with en-- and decoders
$\varepsilon:{\cal B}({\cal M})\rightarrow\widetilde{\cal A}$ and
$D:{\cal P}({\cal M})\rightarrow\widetilde{\cal A}$.
\par
The above definition, eq.~(\ref{eq:quantum:ID}), should be compared with the
usual definition of a quantum transmission code: there $D$ in
eq.~(\ref{eq:quantum:decoder}) is (the restriction of) a completely
positive, unit preserving map (the adjoint of the decoder channel $\delta$)
\begin{equation}
  \label{eq:blind:decoder}
  \delta^*:{\cal B}({\cal M})\longrightarrow{\cal A}_2^{\otimes n},
                           \text{ such that }D_\tau = \delta^*(\tau).
\end{equation}
Notice that in this case eq.~(\ref{eq:quantum:ID}) implies the familiar
fidelity condition, that for all $\pi$,
$\tr\bigl[ \bigl((\delta T^{\otimes n}\varepsilon)\pi\bigr) \pi \bigr] \geq 1-\lambda/2$.
On the other hand it is implied by
\begin{equation}
  \label{eq:quantum-transmission}
  \forall\pi\quad \left\| \bigl((\delta T^{\otimes n}\varepsilon)\pi\bigr) - \pi \right\|_1
                                                                              \leq \lambda.
\end{equation}
In this sense our concept of quantum--ID code (eqs.~(\ref{eq:quantum:encoder}),
(\ref{eq:quantum:decoder}))
is a ``visible decoder'' version of quantum transmission: to be precise, we
shall call it \emph{ID--visible}.
A decoder of the form eq.~(\ref{eq:blind:decoder}) we call \emph{blind}.
Noticing that a blind decoder allows to perform any measurement, not
only the fidelity--test, on the input state ($\delta^*$ translates POVMs into POVMs),
we are motivated to also define (general) \emph{visible} decoding as a map
\begin{equation}\begin{split}
  \label{eq:visible}
  D:{\rm POVM}({\cal M}) &\longrightarrow {\rm POVM}({\cal A}_2^{\otimes n}) \\
         {\bf M}=(M_y)_y &\longmapsto            D({\bf M})=(M_y')_y,
\end{split}\end{equation}
with the error criterion that for all $\pi\in{\cal S}({\cal M})$,
\begin{equation}
  \label{eq:V:criterion}
   \sum_y \left| \tr(\pi M_y)
                -\tr\bigl[\bigl((T^{\otimes n}\varepsilon)\pi\bigr)M_y'\bigr] \right|
                                                                          \leq \lambda.
\end{equation}
Note that this is implied by eq.~(\ref{eq:quantum-transmission}) and in
turn implies eq.~(\ref{eq:quantum:ID}).
\par
All this motivates us to also distinguish blind and visible \emph{encoders}
as well (see Barnum \emph{et al.}\cite{BFJS} and Bennett \emph{et al.}\cite{rand}
for a discussion of these concepts in quantum source coding and remote state preparation):
\emph{blind} is our definition, eq.~(\ref{eq:quantum:decoder}) above,
\emph{visible} is an encoder map
\begin{equation}
  \label{eq:visible:encoder}
  E:{\cal P}({\cal M})\longrightarrow{\cal S}\bigl({\cal A}_1^{\otimes n}\bigr)
\end{equation}
from the pure states on ${\cal M}$ into the state space of
${\cal A}_1^{\otimes n}$.
\par
The type of encoder/decoder we will indicate by a two--letter code, such
as ``bv'' or ``vV'': the first letter refers to the encoder (b: blind, v: visible),
the second to the decoder (b: blind, V: (general) visible, v: ID--visible).
Our original definition above is a bv--code, and this will be the default,
unless otherwise stated.
\par
We begin our investigation observing that all the $L(n,\lambda)$--functions
exhibit only exponential growth:
\begin{prop}
  \label{prop:L-growth}
  For all $0<\lambda<1$ and $\epsilon,\mu>0$, such that
  $\mu\log\left(\frac{1}{\epsilon}-1\right)>4$,
  \begin{equation*}\begin{array}{ccccccc}
    1+\frac{2}{\epsilon}\log N(n,\lambda/2,\lambda/2+\mu)
            & \geq & L_{vv}(n,\lambda) & \geq & L_{vV}(n,\lambda) & \geq & L_{vb}(n,\lambda) \\
            &      &      \geq         &      &      \geq         &      &      \geq         \\
            &      & L_{bv}(n,\lambda) & \geq & L_{bV}(n,\lambda) & \geq & L_{bb}(n,\lambda).
  \end{array}\end{equation*}
\end{prop}
\begin{beweis}
  Only the first inequality is nontrivial.
  For it, simply concatenate a given $(n,\lambda)$--quantum--ID code
  of maximal dimension $L=L_{vv}(n,\lambda)$ with a pure state classical
  ID code with error probabilities of first and second kind
  $0$ and $\mu$, respectively: this gives an
  $(n,\lambda/2,\lambda/2+\mu)$--ID code
  of cardinality $2^{\lfloor\epsilon L\rfloor}/L$, and observing
  $L_{vv}(n,\lambda) \leq N(n,\lambda/2,\lambda/2+\mu)$
  the claim follows.
\end{beweis}
\par
This means that the definition of the various capacities
($xy\in\{vv,vV,vb,bv,bV,bb\}$) has to be
$$Q_{xy}(T)
  := \inf_{\lambda>0}\liminf_{n\rightarrow\infty}\frac{1}{n}\log L_{xy}(n,\lambda),$$
like ordinary transmission capacity $Q(T)$. $Q_{\rm ID}(T)$ will denote
the default $Q_{bv}(T)$. Note that we adopt the
``pessimistic'' notion of capacity: achievable rates have to be met for
every sufficiently large block length $n$. It is conceivable that
the ``optimistic'' definition, where the rate has to be met only
infinitely often ($\limsup$ in the above formula) is larger, not to
mention the question of the strong converse (allowing fixed positive $\lambda$).
\begin{rem}
  \label{rem:Q:somecases}
  For every $T$, $Q_{bb}(T)=Q(T)$, the transmission capacity. This identity extends to
  $Q_{vb}$ in some cases: $Q_{vb}(\id_{\C^2})=1=Q(\id_{\C^2})$ and
  for entanglement--breaking $T$, $Q_{vb}(T)=0=Q(T)$. Both statements are implied
  by a result in Bennett \emph{et al.}\cite{rand}, Theorem 24 (Appendix C), which
  states that to transmit a state of $n$ qubits visibly, using $n-c$ qubits
  requires $\Omega(2^n)$ classical bits: in the first case there is
  no classical channel available, so $c=0$; in the second case the channel
  consists of a measurement followed by a state preparation,\cite{HSR}
  so it cannot be better than a classical channel of bounded capacity.
  This even gives the strong converse for these capacities ($\lambda<1$).
  \par
  For every $T$, $Q_{vv}(T)\leq C_{\rm ID}(T)$: simply concatenate a vv--quantum--ID
  code with the construction of proposition~\ref{prop:many:pure:states}.
  \par
  For every $T$, $Q_{vV}(T)\leq C(T)$: for the states sent through the channels
  once could take an orthogonal basis of ${\cal M}$, and as the POVM the corresponding
  von Neumann measurement, which yields an $(n,\lambda)$--code for $T$.
  If $T$ is entanglement--breaking, then $Q_{bV}(T)=Q_{vV}(T)=0$: as above the
  channel is not better than a classical channel of finite capacity, but
  for this we can invoke a result of
  Ambainis \emph{et al.},\cite{ASTVW} which implies
  that to classically simulate the statistics of arbitrary
  POVMs on arbitrary states of $n$ qubits requires transmission of
  $2^{\Omega(n)}$ bits.
\end{rem}
\par
The following result is the centre--piece of this paper:
\begin{prop}
  \label{prop:2for1}
  For $0<\lambda<1$, there exists on ${\cal B}(\C^d)$ a quantum--ID
  code of error $\lambda$ and
  $\dim{\cal M}=S=\left\lfloor d^2\frac{(\lambda/100)^4}{4\log(100/\lambda)}\right\rfloor$.
\end{prop}
The proof uses the following slight modification of a result
in Bennett \emph{et al.}, Lemma 3 (Appendices A and B):
\begin{lemma}
  \label{lemma:randomunitaries}
  Let $\psi$ be a pure state, $P$ a projector of rank (at most) $r$
  and let $U\in{\cal U}(d)$ be a random variable,
  distributed according to the Haar measure.
  Then for $\epsilon>0$,
  \begin{equation*}
    \Pr\left\{ \tr(U \psi U^* P) \geq (1+\epsilon)\frac{r}{d} \right\}
                            \leq \exp\left( -r \frac{\epsilon-\ln(1+\epsilon)}{\ln 2}\right).
  \end{equation*}
\end{lemma}
\begin{beweis}
  From Bennett \emph{et al.},\cite{rand} proof of Lemma 3, we take that 
  the probability in question is bounded by
  $\exp\left( -r \frac{2}{\ln 2}\Lambda^*(1+\epsilon) \right)$,
  with the rate function (see Dembo and Zeitouni\cite{Dembo:Zeitouni}
  for definitions) $\Lambda^*$ of the square of a real
  Gaussian distributed random variable of mean $0$ and variance $1$.
  This is calculated in Bennett \emph{et al.},\cite{rand} Lemma 23
  (Appendix A), to
  \begin{equation*}
    \Lambda^*(x)=\begin{cases}
                   \frac{1}{2}\left(x-1-\ln x \right) &      :\   x>0,     \\
                   \infty                             &      :\   x\leq 0,
                 \end{cases}
  \end{equation*}
  which gives the claim.
\end{beweis}
\begin{beweis}[of proposition~\ref{prop:2for1}]
  Pick an $\eta$--net, with $\eta=\lambda/8$, in ${\cal P}(\C^S)$ of cardinality
  $\left(\frac{5}{\eta}\right)^{2S}$, according to lemma~\ref{lemma:net}.
  The encoder will be an isometry (with $a<d$ and $S<ad$)
  $V:\C^S\longrightarrow\C^d\otimes\C^a$,
  followed by the partial trace over the second system $\C^a$:
  $\varepsilon(\pi) = \tr_{\C^a} \bigl(V\pi V^*\bigr)$,
  while the decoder is simply given by $D_\tau=\supp\varepsilon(\widetilde{\tau})$,
  the support projector onto the image of $\widetilde{\tau}$, the nearest
  state to $\tau$ in our $\eta$--net. We will fix $S$ and $a$ later.
  \par
  We shall pick the isometry $V$ randomly (uniformly, i.e.~according
  to the invariant measure), and show that with high probability
  it will yield a code of error $\eta$, at least for states from
  the $\eta$--net. Then, for arbitrary states $\pi,\tau$ and their nearest
  neighbours $\widetilde{\pi},\widetilde{\tau}$ in the net,
  \begin{equation*}\begin{split}
    \bigl| \tr\pi\tau - \tr\varepsilon(\pi)D_\tau \bigr|
         &\leq \bigl| \tr\pi(\tau-\widetilde{\tau}) \bigr|
              +\bigl| \tr(\pi-\widetilde{\pi})\widetilde{\tau} \bigr|              \\
         &\phantom{=}
              +\bigl| \tr(\varepsilon(\pi)-\varepsilon(\widetilde{\pi}))D_\tau \bigr|
              +\bigl| \tr\widetilde{\pi}\widetilde{\tau}
                      -\tr\varepsilon(\widetilde{\pi})D_{\widetilde{\tau}} \bigr|  \\
         &\leq 4\eta=\lambda/2,
  \end{split}\end{equation*}
  using triangle inequality, $D_\tau=D_{\widetilde{\tau}}$ and the nonincrease of
  the trace norm under the partial trace.
  \par
  So, fix an ordered pair $(\tau=\ketbra{\theta},\pi=\ketbra{\phi})$ of states from the
  $\eta$--net, and write
  $\ket{\phi}=\sqrt{\alpha}\ket{\theta}+\sqrt{1-\alpha}\ket{\theta^\perp}$,
  with $\alpha=|\bra{\theta}\phi\rangle|^2$.
  Then we can write, with \emph{independent} random unit vectors $\ket{v},\ket{w}$,
  \begin{align*}
    V\ket{\theta}       &= \ket{v}, \\
    V\ket{\theta^\perp} &= \frac{\ket{w}-\bra{v}w\rangle\ket{v}}%
                                {\|\ket{w}-\bra{v}w\rangle\ket{v}\|_2}.
  \end{align*}
  (Note that the denominator vanishes with probability $0$.)
  Now holding $\ket{v}$ constant for the moment, so that $D_\tau=\supp\tr_a\ketbra{v}$
  is a constant, we have
  \begin{equation*}\begin{split}
    \tr(\varepsilon(\pi)D_\tau) &= \bra{\phi} V^* (D_\tau\otimes\1) V\ket{\phi}            \\
                     &= \left(\Bigl(\sqrt{\alpha}
                                    -\sqrt{1-\alpha}\frac{\bra{w}v\rangle}{t}\Bigr)\bra{v}
                              +\frac{\sqrt{1-\alpha}}{t}\bra{w}\right)                     \\
                     &\phantom{======}
                                  D_\tau\otimes\1\
                        \left(\Bigl(\sqrt{\alpha}
                                    -\sqrt{1-\alpha}\frac{\bra{v}w\rangle}{t}\Bigr)\ket{v}
                              +\frac{\sqrt{1-\alpha}}{t}\ket{w}\right)                     \\
                     &= \left|\sqrt{\alpha}-\sqrt{1-\alpha}\frac{\bra{v}w\rangle}{t}\right|^2
                       +\frac{1-\alpha}{t^2}\bra{w}D_\tau\otimes\1\ket{w}                  \\
                     &\phantom{==}
                       +\Bigl(\sqrt{\alpha}-\sqrt{1-\alpha}\frac{\bra{w}v\rangle}{t}\Bigr)
                        \frac{\sqrt{1-\alpha}}{t}\bra{v}w\rangle                           \\
                     &\phantom{==}
                       +\Bigl(\sqrt{\alpha}-\sqrt{1-\alpha}\frac{\bra{v}w\rangle}{t}\Bigr)
                        \frac{\sqrt{1-\alpha}}{t}\bra{w}v\rangle                           \\
                     &= \alpha + \frac{1-\alpha}{t^2}
                                  \bigl[ \bra{w} D_\tau\otimes\1 \ket{w}
                                          -\langle w \ketbra{v} w \rangle \bigr]
  \end{split}\end{equation*}
  where we have denoted $t:=\|\ket{w}-\bra{v}w\rangle\ket{v}\|_2$ and used that
  $\ketbra{v}\leq D_\tau\otimes\1$. Hence, if
  \begin{equation}
    \label{eq:2conditions}
    \langle w\ketbra{v}w\rangle \leq \bra{w}D_\tau\otimes\1\ket{w}
                                     \stackrel{!}{\leq} \epsilon:=(\eta/2)^2, 
  \end{equation}
  we can conclude that $\bigl| \tr\varepsilon(\pi)D_\tau - \alpha \bigr|\leq \eta$.
  It remains to bound the probability of the event in eq.~(\ref{eq:2conditions}):
  according to lemma~\ref{lemma:randomunitaries},
  putting $a=\lfloor \epsilon d/2 \rfloor$, we have
  \begin{equation*}
    \Pr\{ \bra{w}D_\tau\otimes\1\ket{w} > \epsilon \}
                                   \leq \exp\left(-a^2\frac{1-\ln 2}{\ln 2}\right).
  \end{equation*}
  Note that $D_\tau$ can have rank $a$ at most; we will assume $a\geq 1$ from now,
  and will see at the end that the case where $a$ would be zero is trivial.
  Putting this and eq.~(\ref{eq:2conditions}) together with the union bound, this gives us
  \begin{equation*}\begin{split}
    \Pr&\Bigl\{\exists\pi,\tau\text{ from the }\eta\text{--net }
               \bigl| \tr\varepsilon(\pi)D_\tau - \tr\pi\tau \bigr| > \eta \Bigr\}   \\
       &\phantom{==============}
                   \leq \left(\frac{5}{\eta}\right)^{4S}\!\!\!
                         \exp\left( -\frac{1-\ln 2}{\ln 2}
                                     \left\lfloor\frac{\epsilon d}{2}\right\rfloor^2 \right),\\
       &\phantom{==============}
        \leq\left(\frac{40}{\lambda}\right)^{4S}\!\!\!
                         \exp\left( -\frac{1-\ln 2}{16\ln 2} (\lambda/16)^4 d^2 \right)
  \end{split}\end{equation*}
  which is smaller than $1$ by our choice of parameters:
  $\eta=\lambda/8$, $\epsilon=(\eta/2)^2$,
  $S=\left\lfloor d^2\frac{(\lambda/100)^4}{4\log(100/\lambda)} \right\rfloor$.
  That is, except when $S=0$ (which is implied by $a=0$), in which case the
  proposition is trivial.
\end{beweis}
\begin{thm}
  \label{thm:2for1}
  $Q_{bv}(\id_{\C^2})=Q_{vv}(\id_{\C^2})=2$, and in fact the strong converse holds:
  for all $0<\lambda<1$,
  $$\lim_{n\rightarrow\infty}\frac{1}{n}\log L_{bv}(n,\lambda)=
    \lim_{n\rightarrow\infty}\frac{1}{n}\log L_{vv}(n,\lambda)=2.$$
\end{thm}
\begin{beweis}
  This follows immediately by invoking proposition~\ref{prop:2for1}
  for the direct part on $d=2^n$ dimensions,
  and remark~\ref{rem:Q:somecases} for the converse.
  Looking at proposition~\ref{prop:L-growth} and the
  strong converse for $N(n,\lambda_1,\lambda_2)$
  (proposition~\ref{prop:optimality}) yields the strong converse.
\end{beweis}
\begin{cor}
  \label{cor:v-vs-V}
  For every channel $T$,
  $$Q_{bv}(T) \geq 2Q_{bV}(T),\quad Q_{vv}(T)\geq 2Q_{vV}(T).$$
\end{cor}
\begin{beweis}
  We begin by observing that by linearity both blind and visible encoders
  can be understood as mappings ${\cal S}({\cal M})\rightarrow{\cal S}({\cal A}_1^{\otimes n})$.
  The claim follows by understanding that one can concatenate a given
  bV-- or vV--code with the construction of proposition~\ref{prop:2for1}:
  this translates a given pure state on ${\cal M}'$
  into a mixed state on ${\cal M}$ which is then processed further by the
  encoder to a state on ${\cal A}_1^{\otimes n}$;
  it translates a fidelity--test on ${\cal M}'$ into a binary
  POVM on ${\cal M}$ which the visible decoder turns into a POVM on
  ${\cal A}_2^{\otimes n}$.
\end{beweis}
\par
Let us summarise what we can say about two special channels,
$\id_{\C^2}$ (a noiseless qubit) and $\id_2^c$
(the completely dephasing channel, i.e.~a noiseless classical bit):
\par\medskip\noindent
\begin{tabular}{r}
  $T=\id_{\C^2}$: \\
  $C=1$           \\
  $C_{\rm ID}=2$
\end{tabular}
$Q_{xy}\!=\!$
\begin{tabular}{c||c|c|c}
  $x\backslash y$ & b & V & v \\
  \hline\hline
                b & 1 & 1 & 2 \\
  \hline
                v & 1 & 1 & 2
\end{tabular}
\hfill
\begin{tabular}{r}
  $T=\id_2^c$: \\
  $C=1$        \\
  $C_{\rm ID}=1$
\end{tabular}
$Q_{xy}\!\leq\!$
\begin{tabular}{c||c|c|c}
  $x\backslash y$ & b & V   & v \\
  \hline\hline
                b & 0 & $0$ & $1$ \\
  \hline
                v & 0 & $0$ & $1$
\end{tabular}

\section{Conclusions}
\label{sec:conclusion}
Consideration of identification problems via quantum channels has
yielded surprising results:
%
%after showing that \emph{simultaneous}
%ID coding via quantum channels results in a capacity identical to
%the transmission capacity given by the HSW--theorem (at least if one is content
%with a weak converse),
%
we proved that the unrestricted identification capacity can be larger than the
transmission capacity by explicitly computing it for
hybrid quantum memories. These findings are in marked contrast
to the classical situation where the identification capacity shows
a strong connection to the capacity ${\rm CR}(T)$ of the channel to
create common randomness between sender and receiver.\cite{AC,AD2}
(It is possible to find classical channels with $C_{\rm ID}(T)\neq {\rm CR}(T)$
if one allows exponentially growing alphabets\cite{ahlswede:GIT}
or memory.\cite{Kleinewaechter})
Indeed, using the Holevo information bound\cite{Holevo73}
it is not difficult to show that the common randomness
capacity ${\rm CR}(T)$ is at most $C(T)$; since transmission can always be
used to set up correlation, the other inequality is trivial,
so ${\rm CR}(T)=C(T)$
(see also a forthcoming work by the author).
\par
This increase (by a factor of $2$ for the noiseless
qubit channel) may be connected to dense coding, i.e.,
entanglement--assisted transmission.\cite{BSST}
If so, we would learn that the technical condition in
L\"ober's theorem~\ref{thm:simID} on alphabet size is
necessary; also, that something with the conjecture in
remark~\ref{rem:simID} is wrong --- note that
entanglement increases the capacity
of typical qc--channels (e.g.~from $0.585$ to $1$ for the trine
measurement\cite{Holevo:info}), and that ID coding for qc--channels
is automatically simultaneous due to the built--in measurement.
If not related to dense coding, there is another possible,
geometric, explanation for the effect: it may ultimately have to do
with the geometry of state space,
where the dimension of the manifold of mixed states is
roughly the square of the dimension of the pure state manifold.
\par
Then we went further and considered identification of quantum states:
also there we showed that the capacity can exceed the quantum
transmission capacity, and the same factor $2$ occurs for the noiseless
qubit.
Our construction may well be of independent interest, as it
essentially consists of a ``random noisy channel''
$$R_s^{t(u)}:{\cal B}(\C^s)\stackrel{V\cdot V^*}{\longrightarrow} {\cal B}(\C^t\otimes\C^u)
                        \stackrel{\tr_{\C^u}}{\longrightarrow} {\cal B}(\C^t),$$
with an (isotropic) random isometry $V$ and $s\leq tu$.
E.g.~for $s=1$ we recover the generation of random mixed states
by partial trace from a random pure state (for a recent work see
\.{Z}yczkowski and Sommers\cite{ZS}) --- one can rewrite
our construction of a classical ID code
(proposition~\ref{prop:many:mixed:states})
in these terms, with $t=O(u\log u)$;
for $u=1$ we are selecting a random embedding which is
used in proofs of the quantum channel coding theorem.
\par\medskip
We had to leave many open questions: is
$C_{\rm sim-ID} \neq C_{\rm ID}$ and what is $C_{\rm ID}(T)$ in general?
Could it be equal to the lower bound given in theorem~\ref{thm:2Q+C}?
What is $Q_{\rm ID}(T)$ in general --- in particular, can it
be larger than $2Q(T)$? We conjecture that one needs quantum
transmission to have $Q_{\rm ID}(T)\neq 0$ and in particular that
$Q_{\rm ID}(\id_{{\cal A}}) = Q_{vv}(\id_{{\cal A}})
                            = \max\{ 2\log d_i:1\leq i\leq r\}$
for the hybrid quantum memory
${\cal A}=\bigoplus_{i=1}^r {\cal B}(\C^{d_i})$.
Note that this would also imply that (bounded)
common randomness between the users
does not increase the quantum identification capacities, unlike
the classical case\cite{AD2} (see remark~\ref{rem:CR-assistance}):
this is certainly true for $Q_{bb}$ (and likely for $Q_{vb}$,
see Bennett \emph{et al.},\cite{rand} Theorem 24 (Appendix C)
for combinations of noiseless quantum and classical channels) because
we can phrase the error condition as an average pure state fidelity,
and this can always be achieved without randomisation.
Similar questions arise in
connection with the other visible coding capacities
that we defined: for example, it seems that visible \emph{encoding}
and bounded capacity classical side channels
cannot increase our quantum capacities (unbounded classical side channels
will certainly trivialise the visible encoder variants).
How could one prove this in general? A possible
strategy that would solve this and relates to the
above question about $Q_{\rm ID}(T) \stackrel{{\rm ?}}{=}2Q(T)$
could be the construction of a simulation of $T$
for visible transmission, using $Q(T)$ qubits and some finite rate
of classical bits. Entanglement binding channels however might pose
a difficulty here (note that they are the reason that a simulation
can only possibly exist for visibly given input states).
\par
The role of the general visible \emph{decoding} (``V'') is not yet very
clear, but extremely challenging: we conjecture that this model
is as hard as usual transmission but it seems one needs methods beyond
those of Schumacher \emph{et al.}\cite{schumacher96} to prove any bound.
One such approach could be based on a further investigation of the state
space geometric relations in ID coding: to begin with, classical
identification is an almost--isometry (under statistical distance)
from the vertices of an $O(2^d)$--probability simplex into a $d$--probability
simplex; quantum fingerprinting maps similarly into the pure
states on $\C^d$ (with trace distance).
Our quantum--ID codes for the noiseless channel are almost--isometries
(under trace distance) from the set of pure states on $\C^{O(d^2)}$ into the
mixed states on $\C^d$; note that here even the dimensions of the two
state spaces as manifolds are comparable! In fact, what is more important,
they have comparable Vapnip--Chervonenkis dimension
(also called \emph{covering dimension}).
For the V-- and b--variants of decoders stricter metric constraints apply,
extending to larger groups of states, not only pairwise distances,
which accounts for the capacities being smaller in these models.

\section*{Acknowledgements}
Thanks to Rudolf Ahlswede, Masahito Hayashi, Patrick Hayden,
Debbie Leung, Keiji Matsumoto and Peter Shor for stimulating
conversations on identification theory, and especially to Peter L\"ober
for discussions about the proper definition of ``quantum identification''.
I am very grateful to Yossef Steinberg for pointing out
a crucial eror in an earlier version of this paper.
%
%I benefited from discussions with Masahito Hayashi on the strong
%converse for qc--channels, and wish to thank
%Keiji Matsumoto for explaining his proof of that fact.
%
\\
The author was supported by the U.K.~Engineering and Physical
Sciences Research Council. Part of this work was carried out 
at the ERATO Quantum Computation and Information
project, Tokyo, during a visit in April 2002.

\end{document}